\documentclass[apj]{emulateapj}

\def\ergs{ergs~s$^{-1}$}
\def\kms{km~s$^{-1}$}

\begin{document}
\title{
       Spectroscopic studies of an ultraluminous supersoft X-ray source in M81
       }

\author{Yu Bai, JiFeng Liu\footnotemark[$\dagger$] and Song Wang }
\affil{Key Laboratory of Optical Astronomy, National Astronomical Observatories, Chinese Academy of Sciences,
       20A Datun Road, Chaoyang Distict, 100012 Beijing, China}
\footnotetext[$\dagger$]{jfliu@nao.cas.cn}

\begin{abstract}

Ultraluminous supersoft X-ray sources (ULS) exhibit supersoft X-ray spectra
with blackbody temperatures below 0.1 keV and bolometric luminosities above
10$^{39}$ \ergs.  
In this Letter we report the first optical spectroscopic observations of a ULS
in M81 using the LRIS spectrograph on the Keck I telescope.
The detected Balmer emission lines show the mean intrinsic velocity dispersion
of 400 $\pm$ 80 \kms,  which are consistent with from an accretion disk.  The
spectral index of the continuum on the blue side is also consistent with the
multi-color disk model.  The H$_{\alpha}$ emission line exhibits a velocity of
$\sim$ 180 km/s relative to the local stellar environment, suggesting this ULS is
possibly a halo system in M81 belonging to an old population.
%an immigrant system or a foreground supersoft X-ray source in our Galaxy.
No significant shift is found for the H$_{\alpha}$ emission line between two
observations separated by four nights.

\end{abstract}

\keywords{galaxies: individual (M81) --- X-rays: binaries}

\section{Introduction}

Ultraluminous supersoft X-ray sources (ULSs)
are pointlike, non-nuclear X-ray sources
having extremely soft spectra with equivalent blackbody temperatures below 0.1 keV
and bolometric luminosities above 10$^{39}$ \ergs.
They are thought to be massive white dwarfs (WDs) burning accreted material on their surface or
intermediate-mass black holes (IMBHs; 10$^{2}-$10$^{4}$ $M_{\odot}$) with sub-Eddington accretion \citep{Liu08}.

\citet{Swartz02} observed the nearby spiral galaxy M81 with \textit{Chandra} ACIS, and
discovered an intriguing ULS in the bulge, R.A. = 09$^h$55$^m$42.2$^s$,
dec. = $+$69{\arcdeg}03{\arcmin}36.5{\arcsec} (J2000.0) (heareafter M81 ULS1).
Its spectrum can be fitted by a blackbody model with a temperature of $\sim$ 70 eV.
The bolometric luminosity calculated with the distance of M81 is $\sim$ 10$^{39}$ \ergs.
The follow-up X-ray study reveals that
its spectrum can be described by either a blackbody for a WD
or a multi-color accretion disk for an IMBH \citep{Liu08b}.

Its optical counterpart was detected by \textit{Hubble Space Telescope} ($HST$; \citealt{Liu08}).
The spectral energy distribution (SED) from broad-bands photometry exhibits a blue and a red component.
The spectral index of the blue component is consistent with a geometrically thin
accretion disk, and the red component could be described by an AGB star.
The SED also shows excessive H$_{\alpha}$ emission
which is probably originated from the accretion disk or surrounding material
photo-ionized by the soft X-ray emission of M81 ULS1, but the band photometry results cannot
provide the width of the emission line known as an indicator of physical process.

The SED measurements, however, suffer from the intrinsic variabilities of ULS1,
since a flux decrease by a factor of $\sim$ 2.4, as \citet{Tao11} presented,
occurred in less than a week in optical wavelength.
Spectral observations, on the other hand, are independent of the variabilities.
The expected Balmer emission lines in the spectrum will enable us
not only to characterize its physical conditions but also to probe its local
environments \citep{Moon11}. This information is essential for us to understand
the nature of M81 ULS1.

In this Letter we report Keck spectroscopic observations of M81 ULS1 and present evidence
for the existence of an accretion disk.
Section~\ref{opt} describes the Keck/LRIS observations and the result.
The discussion is given in Section~\ref{dis}.

\section{Data Analysis and Results} \label{opt}
The observations of M81 ULS1 occurred on 2010 April 13 and 2010 April 17 during its
expected X-ray-low state \citep{Wang15} using the Low Resolution Imaging Spectrograph
(LRIS) on the Keck I 10m telescope.
Three exposures of 1000 s were taken in the first night and two of 1200 s in
the second night.
The mean seeings in $B$ band were 0.6$\arcsec$ and 0.8$\arcsec$ respectively.
The light of the counterpart was masked with a 0.7{\arcsec} wide slit and split
with a beam dichroic to the blue and red sides followed
by a 300 lines mm$^{-1}$ and a 400 lines mm$^{-1}$ grating.

\begin{figure}
   \includegraphics[width=8cm]{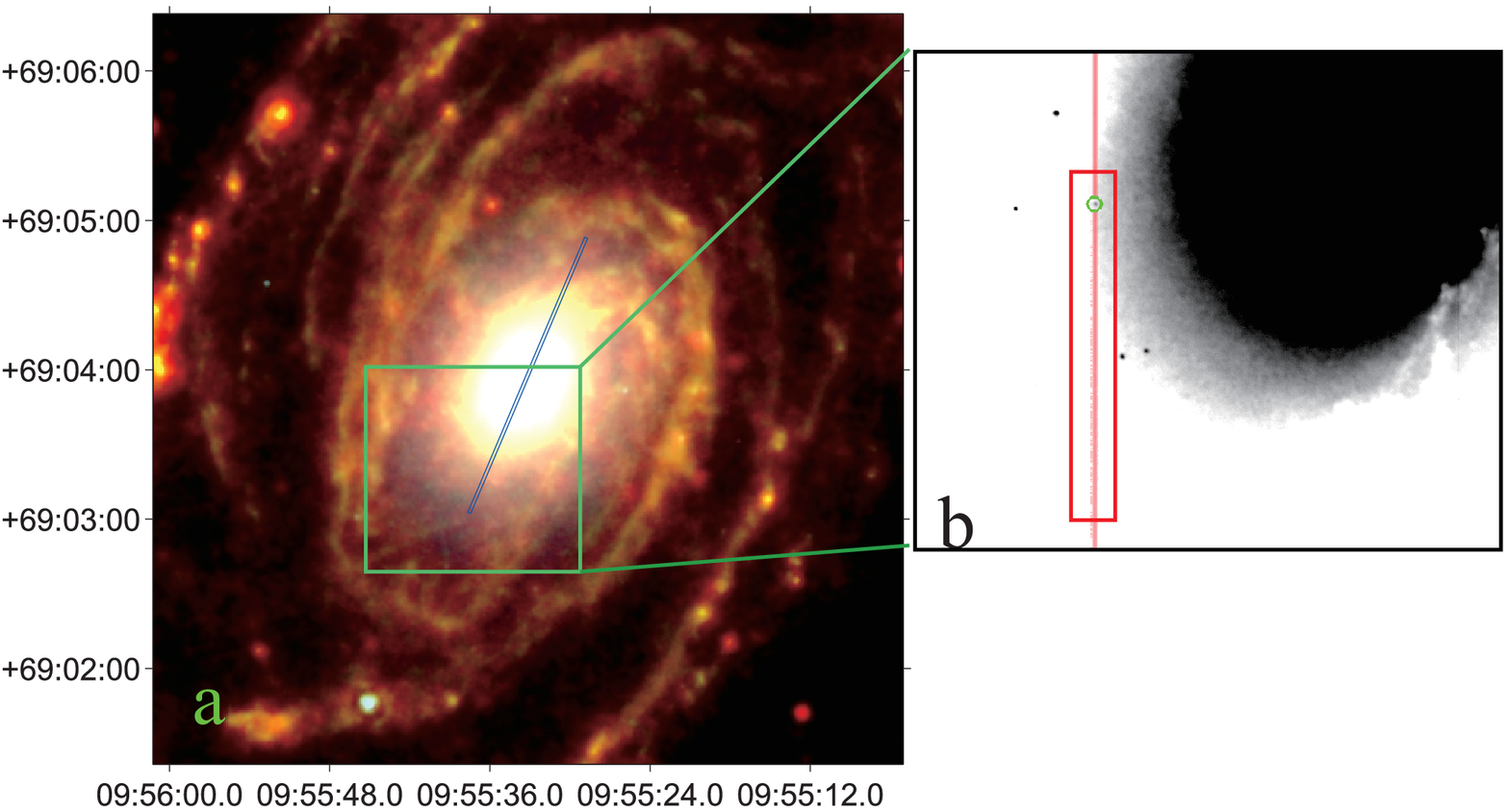}
   \includegraphics[width=8cm]{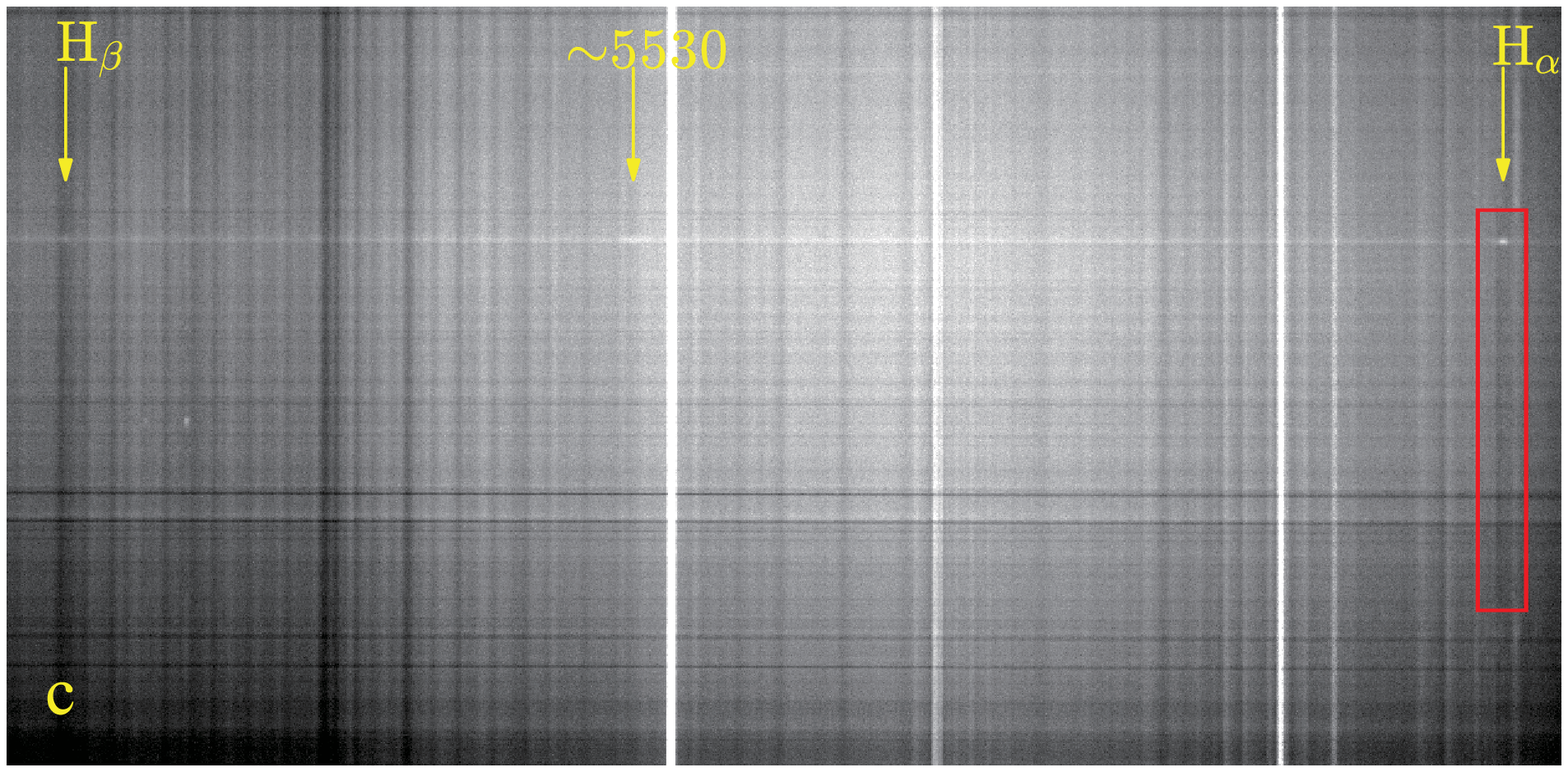}
   \caption{
   \textit{a}: False-color \textit{Spitzer} map of M81 used 5.8 $\mu$m ($blue$), 8.0 $\mu$m ($green$)
               and 24 $\mu$m ($red$).
               The blue line is 2{\arcsec} slit in \citet{Vega01}.
               The light green rectangle presents the sky covered by Keck.
   \textit{b}: The target acquisition image (gray) on the blue side
               and the image of 0.7{\arcsec} wide slit (red) in April 13.
               The green circle gives the position of the optical counterpart in a radius of 1\arcsec.
   \textit{c}: A section of the 2-D dispersed image in April 13. The yellow arrows mark the locations
               of H$_\beta$, H$_\alpha$ and the unknown emission line around 5530 {\AA}.
               The red rectangle marks the H$_{\alpha}$ absorption line in a region
               corresponding to the red rectangle in \textit{b}.
   }
   \label{position}
\end{figure}

The spectra were reduced in a standard way with IRAF.
First of all, raw FITS files were bias-subtracted,
flat-corrected, and combined.
On the blue side of the spectra, the position of the optical counterpart along the slit
was verified by comparing its position in the target acquisition image
(Figure \ref{position}\textit{b}) with an \textit{HST} ACS F606W image \citep{Liu08}.
On the red side, since the counterpart was not obvious
along the slit, an offset between the counterpart and
the WD (PG$1708+602$) was used to verify the position of the counterpart.

Subsequently, raw spectra of the counterpart were extracted with an aperture size of 1\arcsec.
The wavelength calibration was then carried out based on the line lists given in the
manual of Keck\footnote{http://www2.keck.hawaii.edu/inst/lris/.}. The precision
of the calibration is 0.2 {\AA} which is obtained from RMS (root mean square)
of arc-lamp fitting.

\begin{figure}
\center
   \includegraphics[width=\columnwidth]{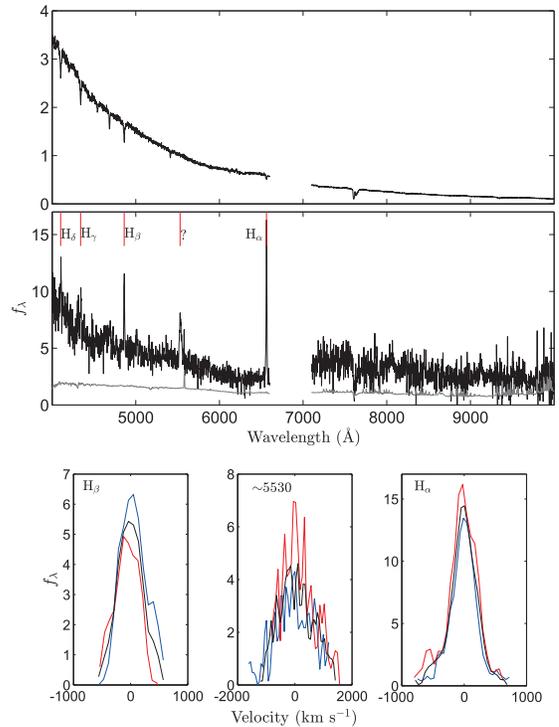}
   \caption{\textit{Upper panel}: Observational result of PG$1708+602$.
            The fluxes are in the unit of 10$^{-14}$ erg s$^{-1}$ cm$^{-2}$ \AA$^{-1}$.
            \textit{Middle panel}: The spectrum of optical counterpart.
            The integrated spectra of counterpart are smoothed to illustrate features clearly.
            Grey lines plot error spectra in 1 $\sigma$ and the positions of hydrogen lines are given in red.
            The fluxes are in the unit of 10$^{-18}$ erg s$^{-1}$ cm$^{-2}$ \AA$^{-1}$ respectively.
            \textit{Lower panel}: Emission lines of H$_{\beta}$, H$_{\alpha}$ and the emission line
            around 5530 {\AA} after the subtraction of the continuum. Red lines stand for
            emission lines obtained in 2010 April 13, blue for
            lines obtained in 2010 April 17 and black for combined
            emission lines.
            The fluxes are in the unit of 10$^{-18}$ erg s$^{-1}$ cm$^{-2}$ \AA$^{-1}$.
            }
   \label{spectra}
\end{figure}

Finally, PG$1708+602$ was used as the standard star to calibrate the flux
by applying the standard flux given by \citet{Massey90}.
  Since the standard flux of PG$1708+602$ covered the wavelength from 3126 {\AA} to
8004 {\AA}, we had to extend the tabulated values to 10000 {\AA} for the calibration
on the red side (the upper panel in Figure \ref{spectra}).

\begin{figure}
\center
   \includegraphics[width=\columnwidth]{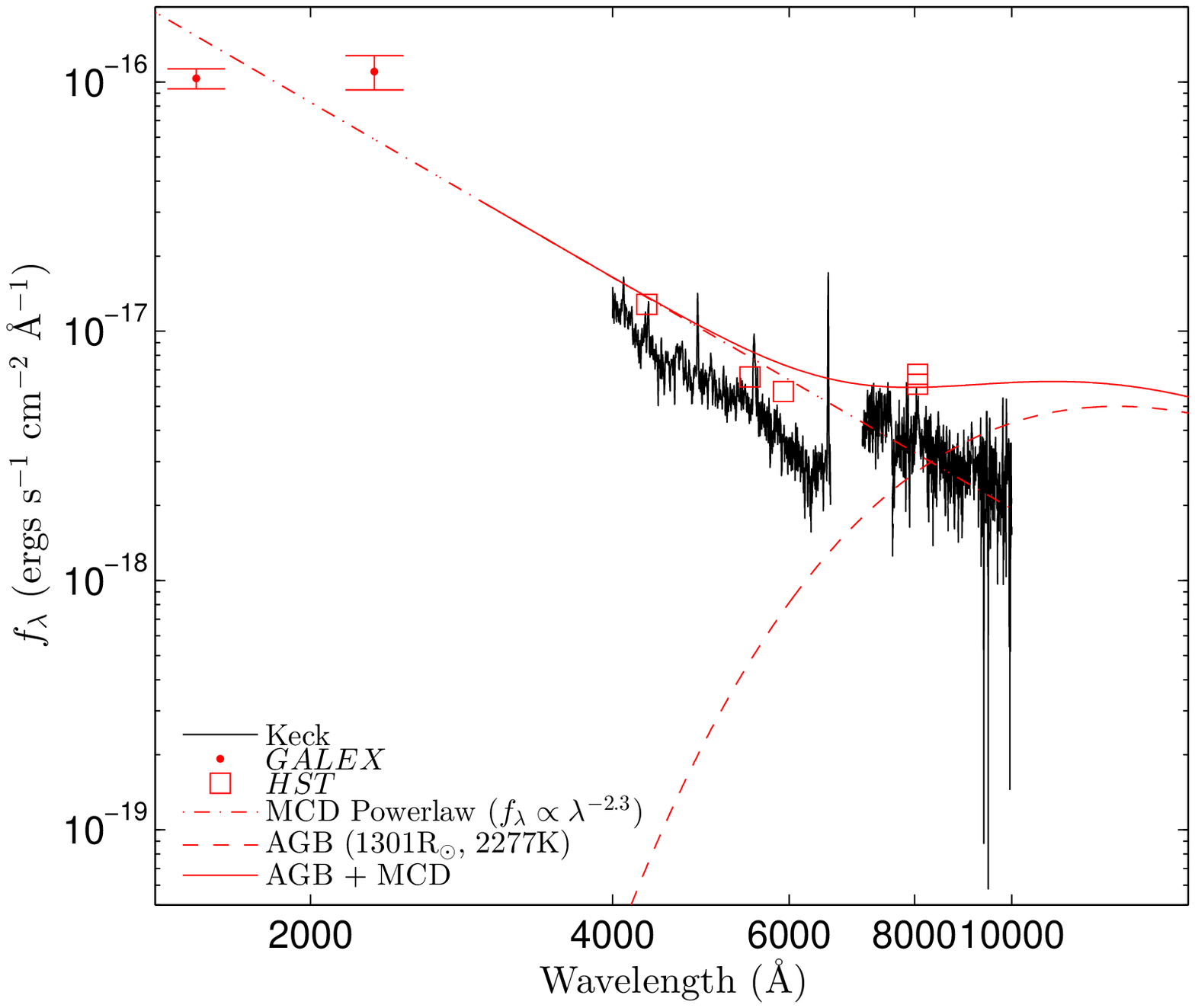}
   \caption{ SED of M81 ULS1.
           The blue and red components suggested by \citet{Liu08} are plotted in red lines.
           Red points are photometric results of \textit{GALEX} and squares are those of \textit{HST}
           obtained from \citet{Liu08}.
            For correction of Galactic extinction, we have adopted the Galactic color excesses, $E(B-V)$=0.08,
            given by \citet{Schlegel98}, and the parameterization
            of the Galactic extinction law given by \citet{Cardelli89} with an extinction ratio
            $R_{V}=3.1$, and the conversion factors for the FUV and NUV
            are $A_{FUV}=7.9E(B-V)$ and $A_{NUV}=8.0E(B-V)$.
           }
   \label{sed}
\end{figure}

\begin{deluxetable*}{ccccccc}
\tablecaption{Line fitting results \label{line}}
\tablehead{
\colhead{}&\multicolumn{3}{c}{Center (\AA)}&\multicolumn{3}{c}{FWHM (\AA)} \\
\colhead{}&\colhead{4.13}&\colhead{4.17}&\colhead{Combined}&\colhead{4.13}&\colhead{4.17}&\colhead{Combined}
}
\startdata
H$_{\beta}$   &4861.4$\pm$1.1&4861.2$\pm$0.3&4861.3$\pm$0.6 & 8.8$\pm$2.7  &8.2$\pm$0.7   &8.6$\pm$1.3  \\
$\sim$5530\AA &5532.3$\pm$1.5&5543.1$\pm$1.6&5536.6$\pm$1.6 & 32.5$\pm$3.5  &32.9$\pm$4.0  &32.2$\pm$3.8 \\
H$_{\alpha}$  &6562.9$\pm$0.2&6562.8$\pm$0.2&6562.8$\pm$0.2 & 10.8$\pm$0.5  &9.4$\pm$0.3   &10.2$\pm$0.4 \\
H$_{\alpha,ab}$$^{\ast}$    &6558.8$\pm$0.5&6558.9$\pm$0.3
\enddata
\tablecomments{
$^{\ast}$Calculated for the H$_{\alpha}$ absorption as marked by the red rectangle in Fig1c.
}
\end{deluxetable*}

\begin{deluxetable*}{ccccccccc}
\tablecaption{\textit{GALEX} Observations and results for M81 ULS1 \label{result}}
\tablehead{
\colhead{Tile}& \colhead{Band} & \colhead{Effective} & \colhead{Exptime} & \colhead{Min} & \colhead{Max} & \colhead{Inner} & \colhead{Outer} & \colhead{AB Mag.} \\
              &                & \colhead{Wavelength} &  & \colhead{ObsDate} & \colhead{ObsDate} & \colhead{Radius} & \colhead{Radius} &  \\
\colhead{(1)} & \colhead{(2)} & \colhead{(3)} & \colhead{(4)} & \colhead{(5)} & \colhead{(6)} & \colhead{(7)} & \colhead{(8)} & \colhead{(9)} }
\startdata
 GI1\_071001\_M81 & FUV & 1538.6 & 14706.7 & 2006-01-05 & 2007-03-31 & 6.3 & 10.5 & 22.27$\pm$0.10 \\
                  & NUV & 2315.7 & 29421.5 & 2005-01-12 & 2007-03-31 & 7.9 & 13.2 & 21.36$\pm$0.17
\enddata
\tablecomments{
Col. (1): Tile name.
Col. (2): Band name. Col. (3): Effective wavelength in angstrom.
Col. (4): Total exposure time in seconds.
Col. (5): Earliest observation UT date for visits which make up the coadd.
Col. (6): Latest observation UT date.
Col. (7): Inner radius in arcsecond for signal integration \citet{Morrissey07}.
Col. (8): Outer radius in arcsecond for background substraction.
Col. (9): Magnitudes in AB system without correction of Galactic extinction. }

\end{deluxetable*}

\subsection{Balmer Emission Lines} \label{bal}

As shown in Figure \ref{spectra}, the Balmer emission  lines are notable
features in the spectra.  Here we used $\chi^2$ minimization to fit the centers
and FWHMs (full width at half-maximum) of H$_{\beta}$, H$_\alpha$ and another
notable emission line around 5530{\AA}. The results with 1$\sigma$ errors are
listed in Table~\ref{line}.  The mean FWHM of H$_{\beta}$ and H$_{\alpha}$
emission lines derived from the fitting is 490 $\pm$ 80 \kms, which is
significantly larger than the spectral resolution, 280 $\pm$ 10 \kms measured
from Hg I $\lambda$5461 in the arc lamp spectrum. The intrinsic dispersion is
400 $\pm$ 80 \kms for the Balmer emission lines.
%
%Here for comparison the radial velocity of PG1708+602 is 5$\pm$8 \kms calculated from H$_{\beta}$
%absorption line, which is consistent with zero within the error.
%
The radial velocity of ULS1 is consistent with zero (an average of 2 $\pm$ 23
\kms) within the precision of wavelength calibration, which implies that M81
ULS1 is unlikely a distant AGN with a large receding velocity.

The Balmer emission lines are mainly seen in spectra of H II regions, planetary
nebulae and accretion disks around compact objects. The line dispersions for H
II regions and planetary nebulae should be comparable to instrumental
dispersions \citep{Fang13,Nicholls14}, while for accretion disks the line
dispersions range from a few hundreds to thousands of kilo meters per second.
The detected Balmer emission lines of M81 ULS1 are significantly broader than
the instrumental dispersion, and are consistent with from an accretion disk
around a compact object.

The observations in two nights enable us to measure the shift of H$_{\alpha}$
related to the binary motion.  We used the technology of the cross-correlation
phasor to estimate the relative velocity.  The technology is based on the
correlation in the wave-number space and it can provide more information than
the normal cross-correlation, such as the significant level (see
\citealt{Misra2010} for detail). The shifts of H$_{\alpha}$ and sky emission
lines are measured in the pixel space to avoid the uncertainty of the
wavelength calibration. The relative shift of H$_{\alpha}$ to the sky line is
0.43$\pm$0.19 pixel, corresponding to 0.61$\pm$0.38 {\AA} toward short
wavelength.  Then we used $\chi$-square minimization to check the relative
shift and the result is 0.42$\pm$0.36 pixel.  No shift is found for
H$_{\alpha}$ emission line at high significant level in the two nights
observations with a span of 4 nights.

\subsection{Balmer absorption lines}

H$_\alpha$ absorption features are detected in a specific region along the slit
as illustrated in Figure \ref{position}\textit{c}.
The regions with absorption features are located in the same area of the sky
but different parts of the CCD in two nights observations, so these absorption
features are unlikely artifacts of the CCD.  The same features were also found
near H$_{\beta}$, but the signal is too low to derive reliable velocity.

In order to investigate the association between the Balmer absorption lines and
the environment of M81 ULS1, false-color maps were constructed in optical
(\textit{HST}), UV (\textit{GALEX}) and IR (\textit{Spitzer}) wavelengths.
Only the map of \textit{Spitzer} exhibits obvious structures along the slit
(Figure~\ref{position}\textit{a}).
The region with the absorption features is correlated with the dark area in the
map of 8~$\mu$m, and the absorption features disappear when the 8~$\mu$m
emission is strong.
Note that there is a positive correlation between 8~$\mu$m surface brightness
and H$_{\alpha}$ emission powered by star forming activity \citep{Young14}.
The observed absorption features appear in the non-star-forming regions, and
likely come from the local stellar environments in the M81 bulge.

The absorption features in the background exhibit high velocities, $-$180 $\pm$
25 \kms at 52{\arcsec} from the nucleus.  The stellar kinematics studies
\citep{Vega01} show that the bulge of M81 is rotating and the south-east side
is approaching with velocities around $-$180 \kms at 45{\arcsec} from the
nucleus.  The similar velocity ($-$180 \kms) suggests that the observed
H$_{\alpha}$ absorption lines come from stars in the bulge of M81.
In comparison, the systemic velocity of M81 is $-$34 \kms  (obtained from NED).

The H$_{\alpha}$ emission line ($-2 \pm 23$ \kms) exhibits a receding velocity
of 178 $\pm$ 35 \kms relative to the H$_\alpha$ absorption line from the
stellar environment.  
This suggests that ULS1 is not a system comoving with stars in the bulge of
M81.
Such a relative velocity can come from an object in the halo of M81 projected
onto its bulge, which is receding along the line of sight.
Here we rule out the possibility that the system is located in the Milky Way,
since the column density of neutral hydrogen atoms at the location of ULS1 is
5.4 $\pm$ 0.5 $\times$ 10$^{20}$ cm$^{-2}$ in the Milky Way \citep{Guver09},
smaller than that derived from the X-ray spectrum fitting by a factor of 2
\citep{Liu08b}.

\subsection{SED Construction} \label{sedc}
The spectral observations of M81 ULS1 enable us to obtain SED
uncontaminated by its intrinsic variabilities.
Here we removed all the emission lines and used a power-law function, $f_{\lambda} \propto \lambda^{\beta}$,
to fit the continuum of $\sim$ 4500$-$6000 {\AA}, the high response wavelength range,
in order to estimate the spectral index on the blue side.
The $\beta$ derived from the best fitting is $-$2.36 $\pm$ 0.02,
which corresponds to $\alpha\sim$ 0.36 $\pm$ 0.02 for $f_{\nu} \propto \nu^{\alpha}$.
The continuum on the blue side is consistent with
the $f_{\nu} \propto \nu^{1/3}$ relation expected for the multi-color disk (MCD) model.
On the red side, the spectrum of an AGB star modeled with a temperature of 2277 K and
a radius of 1301 R$_{\odot}$, as suggested by \citet{Liu08}, is presented in Figure \ref{sed}.
The combined SED is not consistent with the red-side spectrum, which suggests that
the AGB model may be not a good explanation of the red component.

As a connection of X-ray and optical wavebands,
the UV emission is important to understand the SED of M81 ULS1.
In order to probe the UV emission of M81 ULS1, we used the archive data of
the \textit{Galaxy Evolution Explorer} (\textit{GALEX}; \citealt{Martin05}).
M81 have been observed with three different tiles, and here we used the tile with
exposure time over 10$^{5}$ s to do accurate photometry.
All the sub-exposures of the tile were taken during the expected low state of
M81 ULS1 \citep{Liu08b,Wang15}.
The photometric results are listed in Table~\ref{result} and also shown in Figure \ref{sed}.
Although with dispersions, the UV fluxes are lying along the MCD powerlaw.

\section{Discussion} \label{dis}

In this letter we report the first optical spectroscopic confirmation of an
accretion disk around a ULS.
The broad Balmer emission lines of M81 ULS1 revealed by the Keck/LRIS
observations are consistent with from an accretion disk around a compact
object. 
The spectral index of the continuum on the blue side is also consistent with
the $f_{\nu} \propto \nu^{1/3}$ relation expected by MCD model. 
 The velocity of the H$_{\alpha}$ emission line relative to its local stellar
environments suggests that M81 ULS1 is possibly a halo system in M81 belonging
to an old population.
Careful analysis shows no significant shift of H$_{\alpha}$ emission line with
a span of 4 days. This might suggest a small inclination angle or a small
velocity of the compact object. The latter can come from a long orbital period,
or a very massive compact object such as an IMBH. However, the exact nature of
the compact object is still unknown without monitoring of the long-term motion
of the system.

Besides emission lines of Balmer series, an emission line arises around 5530 {\AA}
on the blue side with an FWHM of 1700 $\pm$ 200 \kms, larger than that of H$_{\alpha}$
by a factor of 4 (the lower panel in Figure \ref{spectra}).
All exposures obtained in two nights show the same emission feature and similar
velocity dispersions, so it is unlikely an artifact or cosmic rays.
This emission line is too broad to be a nebular line, and it is probably not an
Fe II emission line because Fe$^{+}$ ion emits through a huge number of multiplets
scattered across the blue side of the spectrum \citep{Baldwin04,Shapovalova12}.
Since the relative shift is 10 {\AA} $\pm$ 2 {\AA} between two nights observations,
the emission line is unlikely related to the accretion disk.
It is reported that N II $\lambda$5530 and $\lambda$5535 emission lines have been
detected for Ae/Be stars \citep{Mathew10,Mathew11}. However, no other features are
found to support the existence of an Ae/Be star, such as Fe II or O I emission lines.
More spectra with high signal-to-noise ratio are needed to draw a conclusive result.

The spectrum of M81 ULS1 shows a gap between the red and blue sides due to the low
response on the edges of CCDs, and signal-to-noise ratio on the red side is very low.
Further deeper spectroscopic observations with the coverage of $\sim$ 5000$-$9000 {\AA}
could present the complete spectrum with high signal-to-noise ratio on the red side,
which will enable us to characterize the secondary.

\begin{acknowledgements}

The authors thank Prof. Bregman for his constructive comments on the manuscript.
Some of the data presented in this paper were obtained from the Mikulski
Archive for Space Telescopes (MAST).  The authors acknowledge supports from
National Science Foundation of China under grants NSFC-11273028 and
NSFC-11333004, and support from National Astronomical Observatories, Chinese
Academy of Sciences under the Young Researcher Grant.

\end{acknowledgements}

%\newpage

\end{document}